\documentclass{PoS}

\usepackage{amsmath, amssymb, multirow}
\usepackage{setspace}

\newcommand{\hess}{H.E.S.S.}
\newcommand{\gr}{$\gamma$-ray}
\newcommand{\grs}{$\gamma$-rays}

\title{Simultaneous H.E.S.S. and RXTE observations of the microquasars GRS 1915+105, Circinus X-1 and V4641 Sgr}

\ShortTitle{F. Sch\"ussler for the H.E.S.S. Collaboration: Microquasar observations}

\author{\speaker{F. Sch\"ussler}\\
Commissariat à l'énergie atomique et aux énergies alternatives / Irfu, Saclay, France \\        E-mail: \email{fabian.schussler@cea.fr}}
        
\author{P. Bordas\\ Max-Planck-Institut für Kernphysik, Heidelberg, Germany\\ E-mail: \email{pol.bordas@mpi-hd.mpg.de}}
\author{P.M. Chadwick \\ Durham University, Durham, UK\\  E-mail: \email{p.m.chadwick@durham.ac.uk}}
\author{H. Dickinson \\ Iowa State University, Ames, USA\\ E-mail: \email{hughd@iastate.edu}}
\author{J.-P. Ernenwein \\ Centre de physique des particules de Marseille, Marseille, France \\ E-mail: \email{ernenwein@cppm.in2p3.fr}}        
\author{on behalf of the \hess\ Collaboration\\}
\abstract{Microquasars, Galactic binary systems showing extended and variable radio emission, are potential gamma-ray emitters. Indications of gamma-ray transient episodes have been reported in at least two systems, Cyg X-1 and Cyg X-3. The identification of additional gamma-ray emitting microquasars is key for a better understanding of these systems.\\
Very-high-energy gamma-ray emission from microquasars has been predicted to happen during periods of transient outbursts potentially connected with the formation of a jet-like outflow. Contemporaneous observations using the H.E.S.S. telescope array and the \emph{RXTE} satellite were obtained on three microquasars: GRS 1915+105, Circinus X-1 and V4641 Sgr with the aim of detecting a broadband flaring event in the very-high-energy gamma-ray and X-ray bands. We report here on the analysis of these data for each system, including a detailed X-ray analysis assessing the location of the sources in a hardness-intensity diagram during the observations. Finally we discuss the derived upper limits on their very-high-energy gamma-ray flux.\\
The analyses presented here will soon be the subject of a detailed publication of the \hess\ Collaboration~\cite{HESSmQuasars2015}.

}

\FullConference{The 34th International Cosmic Ray Conference,\\
		30 July- 6 August, 2015\\
		The Hague, The Netherlands}

\begin{document}
\section{Introduction}
Microquasars, Galactic X-ray binaries that exhibit spatially resolved, extended radio emission, are thought to be structurally similar with the quasar family of active galactic nuclei (AGN). Although the spatial and temporal scales of AGNs and microquasars are substantially different, both object classes are believed to comprise a compact central object embedded in a flow of accreting material, and both exhibit relativistic, collimated jets, which in AGNs are known to be regions of particle acceleration and non-thermal emission. The resulting radiation spectrum can extend from radio wavelengths into the very-high-energy (VHE; $E_{\gamma}>100$ GeV) \gr\ regime. If similar jet-production as well as efficient particle acceleration mechanisms operate in both AGNs and microquasars and assuming appropriate environmental conditions, this might imply the emission of detectable high-energy \gr\ emission from the latter objects. These environmental conditions are defined primarily by the strength of magnetic fields and corresponding synchrotron energy losses, photon field energy densities to be up-scattered through inverse Compton processes to the gamma-ray domain and target particle densities allowing for efficient proton-proton interactions leading to \grs\ through the decay of neutral pions. 

Multi-wavelength monitoring campaigns involving mainly radio telescopes and X-ray satellites have shown that microquasars show transient outbursts, characterised by the ejection of discrete superluminal plasma clouds. These outburst are usually observed at the transition between low and high luminosity X-ray states~\cite{Fender2004} and the internal states of microquasars can be classified using Hardness-Intensity Diagrams (HIDs) displaying the source X-ray intensity against X-ray colour (or hardness)~\cite{HID_Homan2005}. At the transition from the so-called low/hard state to the high/soft states the steady jet associated to the low/hard state is disrupted. These transient ejections, produced once the accretion disk collapses inwards, are more relativistic than the steady low/hard jets~\cite{Fender2004} and therefore might give rise to high-energy emission visible with TeV \gr\ observatories like \hess\ It can be noted the this phenomenological description is extensible also to neutron-stars although in that case jet radio power is lower by a factor 5--30~\cite{Migliari2006}).

Outburst episodes have also been observed in cases in which the source kept in the hard state without transition to the soft state \cite{HID_Homan2005}. For example, the detection (at the $\sim4\sigma$ level) by the MAGIC telescope of the high-mass black-hole binary Cygnus X-1 took place during an enhanced $2-50\,$keV flux low/hard state as observed with the INTEGRAL satellite, the \textit{Swift} BAT and the \emph{RXTE} ASM~\cite{MAGIC_CygX-1_2007, Integral_CygX1}. To date, Cygnus X-1 is the only well established microquasar that has been reported to emit in the VHE \gr\ band. Only one additional object (Cyg X-3) has so far been detected in high-energy \grs\ (>100 MeV) with AGILE and Fermi-LAT~\cite{CygX3MeV}.

Here we report on contemporaneous observations with \hess\ and \emph{RXTE} of the three microquasars V4641~Sgr, GRS~1915+105 and Circinus X-1. Further details of the analysis presented here can be found in~\cite{HESSmQuasars2015}.

\begin{figure*}[!t]
\vspace{-0.3cm}
\centering
\includegraphics[width=0.73\textwidth]{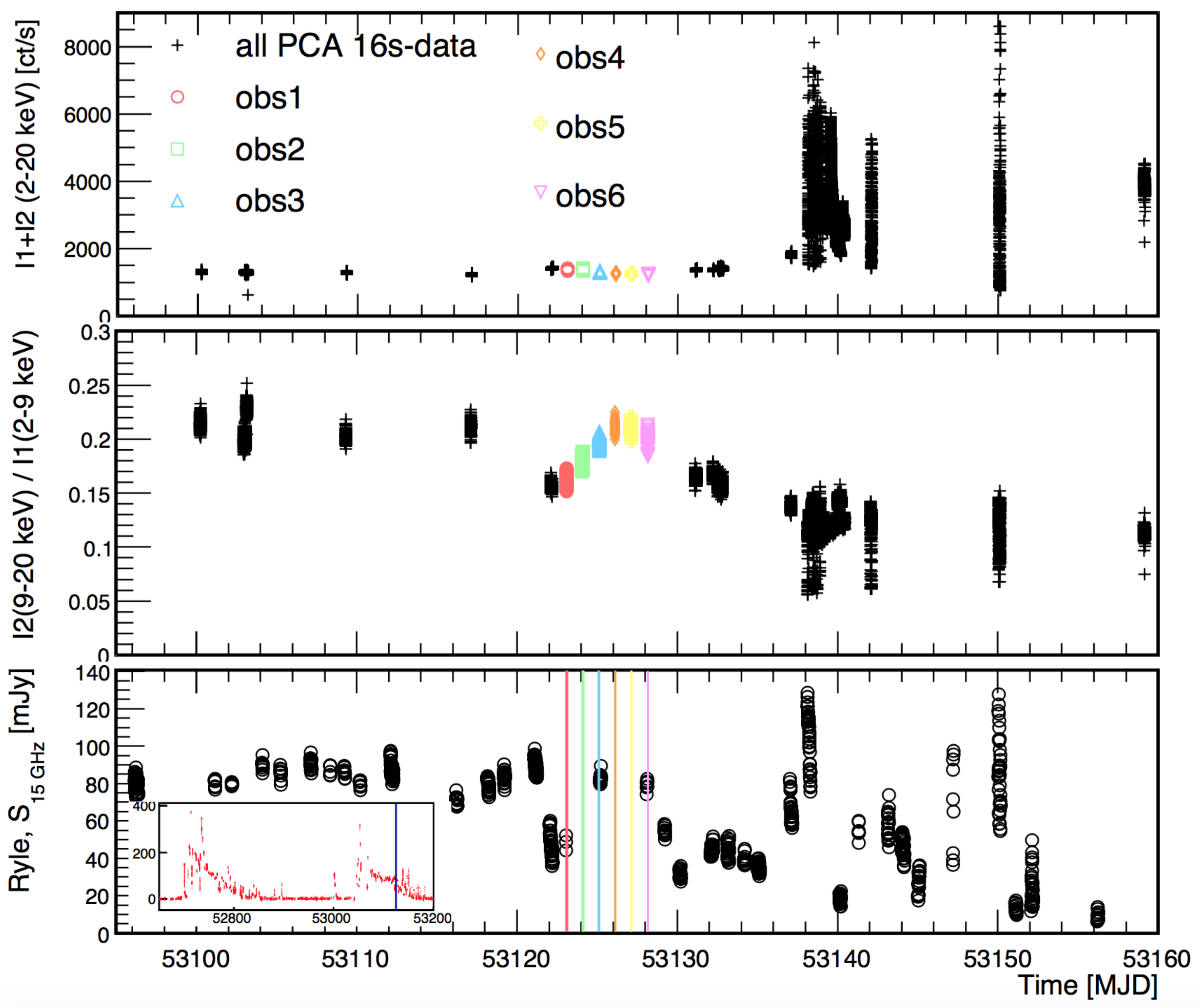}
\caption{Evolution of GRS 1915+105 around the H.E.S.S. observations. Top panel: Intensity of the \textit{\emph{RXTE}} PCA data. Middle panel: evolution of the Hardness derived from \emph{RXTE} PCA data. Bottom panel: contemporaneous  radio data obtained with the Ryle Telescope (15 GHz~\cite{Pooley2006}). A larger, 550 day, time period is depicted in the inset. Data corresponding to \hess\ observations are highlighted with colors.}\label{fig:GRS1915_state}
\end{figure*}

The \hess\ Imaging Atmospheric Cherenkov Telescope array is situated on the Khomas Highland plateau of Namibia (23$^{\circ}16'18''$ South, $16^{\circ}30'00''$ East), at an elevation of 1800 m above sea level. For the analysis presented here, \hess\ observations were carried out using the full original four-telescope array. In this setup \hess\ is sensitive to detect cosmic and gamma-rays in the 100 GeV to 100 TeV energy range and is capable of detecting a Crab-like source close to zenith at the 5$\sigma$ level within $<5$ minutes under good observational conditions~\cite{HESS-Crab2006}. 
The \gr\ analysis followed the standard point-source procedure described in \cite{HESS-Crab2006} using both the \textit{hard} and \textit{standard} event selection cuts. Hard cuts (image size $\ge200$ photoelectrons) will tend to enhance the signal of sources with power-law spectral slopes that are harder in comparison to the dominant cosmic ray background. Standard cuts (image size $\ge80$ photoelectrons) provide less sensitivity in such cases but do allow a lower energy threshold. Upper limits to the VHE $\gamma$-ray flux above the instrumental threshold energy were derived at the 99\% confidence level.\\
The \textit{Rossi X-ray Timing Explorer} (\emph{RXTE}) was a space-based X-ray observatory launched on December 30th, 1995 and decommissioned on January 5th, 2012. The \textit{Proportional Counter Array} (PCA) on board \emph{RXTE} comprised five co-pointing \textit{Proportional Counter Units} (PCUs) which were nominally sensitive in the energy range $\sim2-60$ keV with an energy resolution of $<18\%$ at 6 keV. The \textit{High Energy X-ray-Timing Experiment} (HEXTE) comprised two independent clusters of four \textit{phoswitch} scintillation detectors which were sensitive to photons in the $\sim12-250$ keV energy range and had an energy resolution of $\sim9$ keV at 60 keV. The data presented here were analyzed with the \texttt{FTOOLS 5.3.1} software suite using the data selection criteria regarding elevation, offset, electron contamination and proximity to the South Atlantic Anomaly recommended by the \emph{RXTE} Guest Observer Facility website~\cite{RXTEobserver}. For each observation, the PCA STANDARD2 data were extracted from all available PCUs. HEXTE Archive mode data for both clusters were extracted for all observations following the recommended procedures for time filtering and background estimation. Spectral analyses were carried out using the \texttt{XSPEC 12.6.0} package.

\section{GRS 1915+105}

GRS 1915+105 is a dynamically established black hole binary. Observations in the optical and near infra-red using the Very Large Telescope succeeded in identifying the stellar companion as a low-mass KM III giant~\cite{GRS1915_Greiner}. In a detailed study of the X-ray lightcurves of GRS 1915+105,~\cite{Belloni2000} succeeded in identifying twelve distinct variability classes, internally characterized by the duration and juxtaposition of three separate spectral states. One of these states, defined by a prolonged X-ray plateau, is invariably terminated by flaring activity in the radio, infrared, and X-ray bands~\cite{FenderBelloni2004}. Radio spectra obtained during these end-plateau flaring episodes indicate optically thin synchrotron emission and occasionally show powerful discrete plasma ejections with instantaneous power output reaching $\gtrsim10^{40}$ erg s$^{-1}$ (e.g.~\cite{GRS1915_MirabelRodriguez_Nature}). Modelling the emission from these discrete relativistic ejecta, \cite{GRS195_Aharonian} showed that inverse-Comptonisation of emitted synchrotron photons into the GeV/TeV regime could produce significant and persistent \gr\ fluxes which remain detectable for several days.

For the analysis presented here GRS 1915+105 was observed by \hess\ between April 28th and May 3rd 2004 in response to an apparent decrease in the 15 GHz radio flux which was monitored by the Ryle Telescope during a $\sim50$ day plateau state (G. Pooley, private communication, see lower panel of Fig.~\ref{fig:GRS1915_state}). On the basis of previously observed behaviour, it was thought likely that the observed radio evolution signaled the end of the plateau state and therefore that flaring activity might begin within the subsequent days. As can be seen in the lower panel of Fig.~\ref{fig:GRS1915_state}, the radio flux regained its initial level right after the sudden drop which triggered the observations. The end of the plateau phase and subsequent flaring activity occurred only several days after the observations reported here.

\begin{figure}[!t]
\vspace{-0.3cm}
  \centerline{
   \raisebox{-0.5\height}{\ \includegraphics[width=0.5\linewidth]{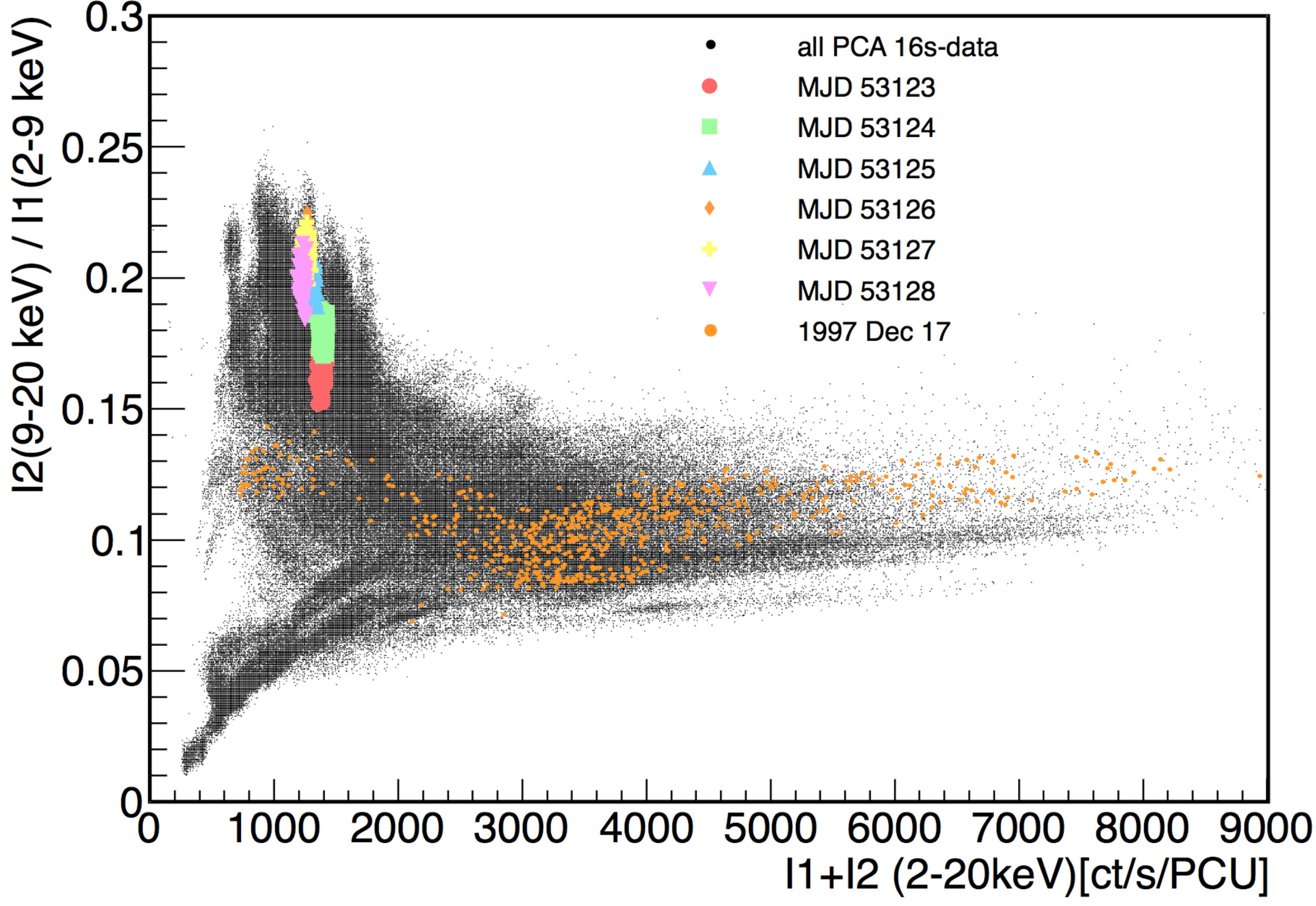}}
  \hfill
  \raisebox{-0.5\height}{\includegraphics[width=0.5\textwidth]{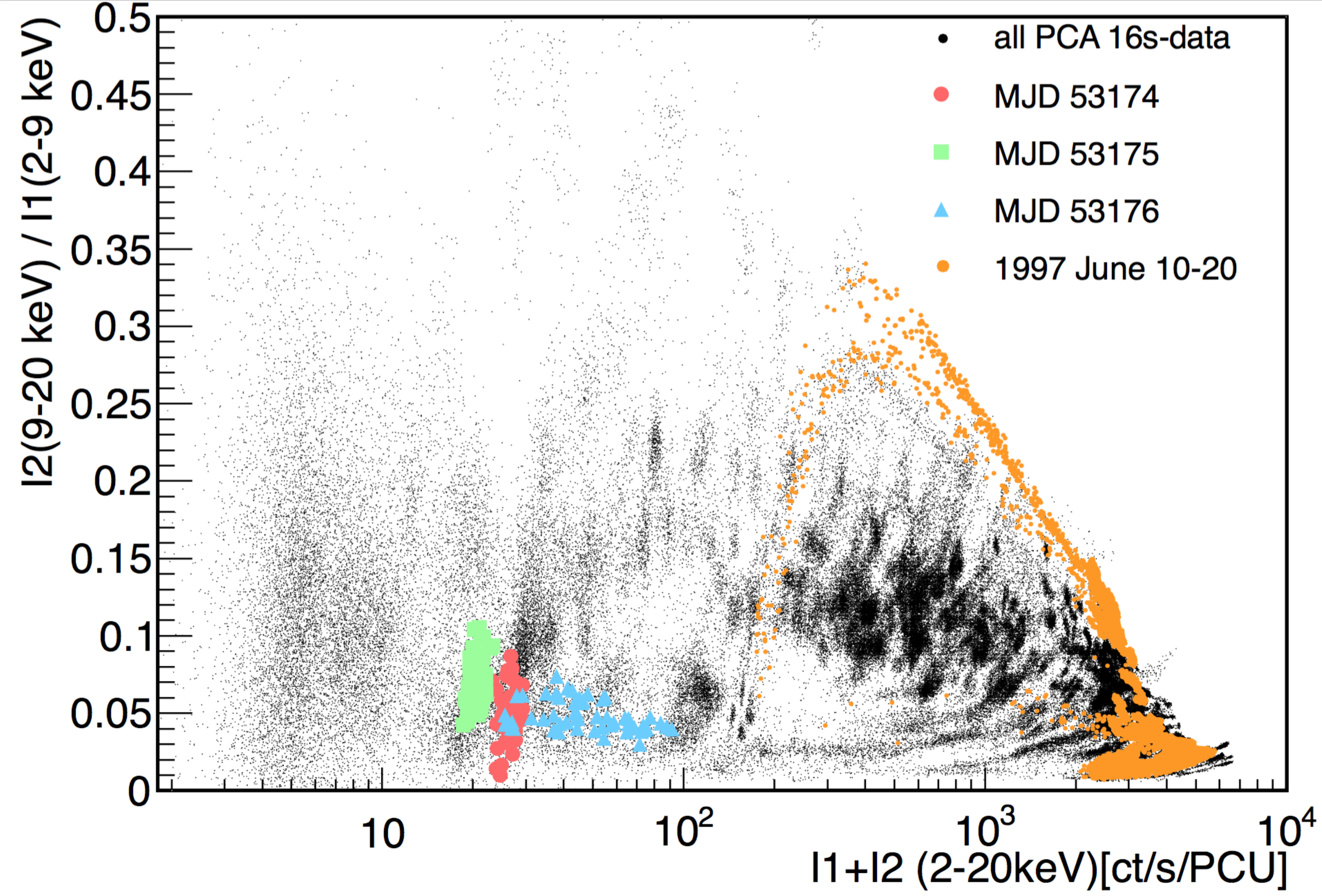}}
  }
   \caption{Hardness-Intensity diagrams of GRS 1915+105 (left) and Circinus X-1 (right) obtained with the \emph{RXTE} PCA data taken during the whole mission of the satellite (black points). Data corresponding to \hess\ observations are highlighted by colored markers. For comparison, periods of known high X-ray activity are also highlighted (e.g.~\cite{Soleri2006} and~\cite{CirX1flares}). Note that for better visibility the axes are scaled differently.}
\label{fig:HIDs}	
 \end{figure}

The \emph{RXTE} observations of GRS 1915+105 comprised six individual pointings, contributing to accumulated PCA and HEXTE livetimes of $7.6\,$ksec and $5176\,$s respectively. Fifteen contemporaneous \hess\ observations were obtained, constituting an overall livetime of 6.9 hours. The X-ray count rate was stable to within $\sim10\%$ during each observation and varied by no more than $\sim20\%$ between observations. Indeed, the long-term \emph{RXTE} PCA lightcurve in the top panel of Fig.~\ref{fig:GRS1915_state} indicates that the \hess\ observation epochs occurred during an extended and relatively faint plateau of the $2-20\,$keV flux. The derived spectra are dominated by a hard, non-thermal component and show temporal stability throughout the period analyzed here. The left plot of Figure~\ref{fig:HIDs} shows the Hardness-Intensity diagrams obtained with the \emph{RXTE} PCA data. The Hardness has been defined as the ratio between the fluxes measured in the bands $\left[9,20\right]$ keV and $\left[2,9\right]$ keV, and the intensity as the sum of the two fluxes, in unit of counts per second. These definition will be used consistently throughout this paper. The time evolution of the hardness of GRS 1915+105 is depicted in the middle panel of Fig.~\ref{fig:GRS1915_state}. The \hess\ observation occurred when the source was close to a Low Hard State (open symbols in Figure \ref{fig:HIDs}), a state in which compact jets are expected to be present. This conclusion seems confirmed by the fact that the observations took place during a radio loud plateau phase. In summary, the combined spectral and temporal analyses indicate a robust association of the contemporaneous \hess\ observation with the \textit{radio-loud} $\chi$ state (e.g.~\cite{Trudolyubov_GRSstates} and the presence of steady, mildly relativistic jets at the time of observation may be confidently inferred.

The contemporaneous \hess\ observations did not yield a significant VHE \gr\ detection. Integral flux upper limits which correspond to the overall \hess\ exposure are listed in Table~\ref{tab:mqs:hess_uls}.

\section{Circinus X-1}\label{sec:cirx1}
Circinus X-1 (hereafter Cir X-1) is a binary system of a low magnetic field  neutron star accompanied either a low-mass/sub-giant companion (implying a high orbital eccentricity $e\sim0.7-0.9$) or a mid-B supergiant (suggesting a more moderate eccentricity $e\sim0.45$). At radio wavelengths, the jets of Cir X-1 display notable structure on arcsecond scales, appearing as a bright core with significant extension along the axial direction of the arcminute jets~\cite{Fender1998_CirJet}. Cir X-1 has also been observed to eject condensations of matter with apparently superluminal velocities $\gtrsim15c$~\cite{FenderRadio}, implying a physical velocity for the ejecta $v>0.998c$ with a maximum angle between the velocity vector and the line of sight $\theta<5^{\circ}$. These results identify Cir X-1 as a microblazar -- a Galactic, small-scale analogue of the blazar class of AGN, several of which are known sources of VHE \grs. 

\begin{figure*}[!t]
\vspace{-0.3cm}
\centering
\includegraphics[width=0.9\textwidth]{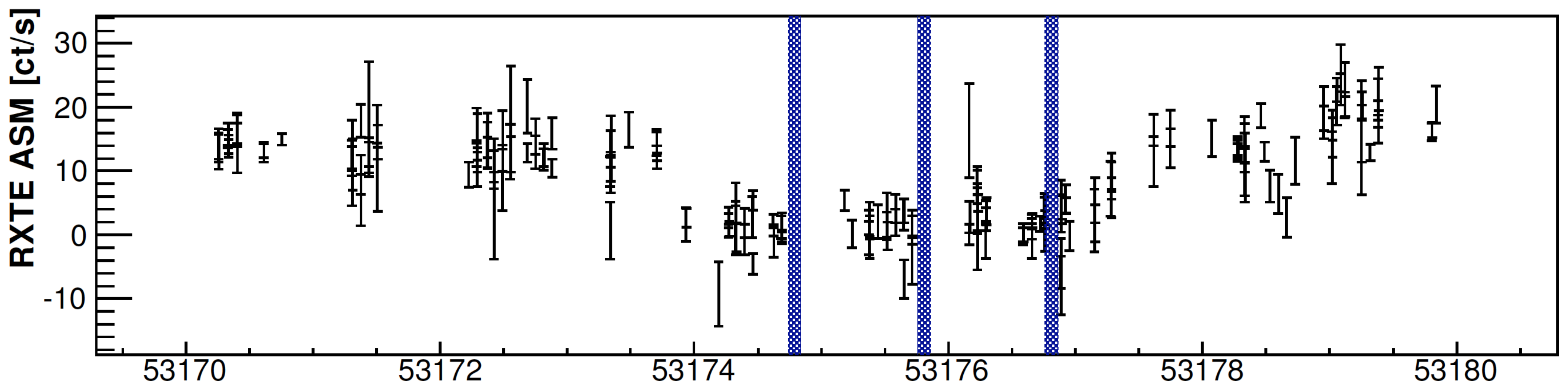}
\caption{\emph{RXTE} ASM lightcurve for Circinus X-1. The blue shaded bands indicate the periods covered by \hess\ observations.}\label{fig:cirx1_lightcurve}
\end{figure*}

The \hess\ observations of Cir X-1 reported here began on June 18th 2004 and were scheduled to coincide with the periastron passage of the binary components. The previous observation of regular radio flares during this orbital interval were thought to provide a good chance of catching Cir X-1 during a period of outburst, with the associated possibility that superluminal ejections might occur.
The \emph{RXTE} observations of Cir X-1 comprised three individual pointings, corresponding to orbital phase intervals $0.0486\le\phi\le0.0498$, $0.1104\le\phi\le0.1112$ and $0.1718\le\phi\le0.1725$, and contributing to an accumulated PCA livetime of $2576\,$s. A dataset comprising 12 contemporaneous \hess\ observations yielded a combined livetime of 5.4 hours. The ASM lightcurve shown in Figure \ref{fig:cirx1_lightcurve} reveals that the \hess\ observations occurred during an extended $\sim4$ day dip in the $2-10\,$keV X-ray flux. Additionally, it should be noted that the observations reported here were obtained during an extremely faint episode in the secular X-ray flux evolution of Cir X-1, as can be seen in the HID depicted in the right panel of Figure~\ref{fig:HIDs}. As a consequence, the measured X-ray fluxes are significantly lower than most others reported for this source. The individual PCA lightcurves obtained during the first two pointings are characterized by a relatively low count rate which remains approximately constant throughout each observation. In marked contrast, the third observation exhibits clear variability with count rates doubling on timescales of 50s. A detailed analysis of the obtained spectra reveals that the observed flux variability is accompanied by marked variations in spectral shape. These can be interpreted as hints towards a strong mass transfer during the periastron passage and subsequent dramatic evolution of the local radiative environment.

\hess\ observations obtained contemporaneously with the \emph{RXTE} pointings yield no detection of VHE \gr\ emission. Derived integral flux upper limits corresponding to the overall \hess\ exposure are listed in Table \ref{tab:mqs:hess_uls}.

\section{V4641 Sgr}\label{sec:v46_res}

V4641 Sgr is the optical designation of the habitually weak X-ray source SAX J1819.3-2525 (XTE J1819-254). Optical spectroscopic measurements (e.g.~\cite{V4641}) strongly suggest a late B/early A-type companion with an effective temperature $T_{\rm{eff}}\approx10500$ K. A derived compact primary mass $8.73 \leq M_{1} \leq 11.7 M_{\odot}$~\cite{V4641} categorises V4641 Sgr as a firm black hole candidate.

Observations of V4641 Sgr with \hess\ were initiated on July 7th 2004 (MJD 53193) in response to the source brightening rapidly in the radio, optical and X-ray bands. The resultant \emph{RXTE} exposure comprised three observations, each contributing to an accumulated PCA livetime of $5\,$ksec. Two pairs of $\sim$30 minute \hess\ observations were obtained contemporaneously with the final two \emph{RXTE} pointings. In total the four separate \hess\ exposures constitute an overall livetime of 1.76 hours.

\begin{figure}[!t]
\vspace{-0.3cm}
  \centerline{
   \raisebox{-0.5\height}{\ \includegraphics[width=0.4\linewidth]{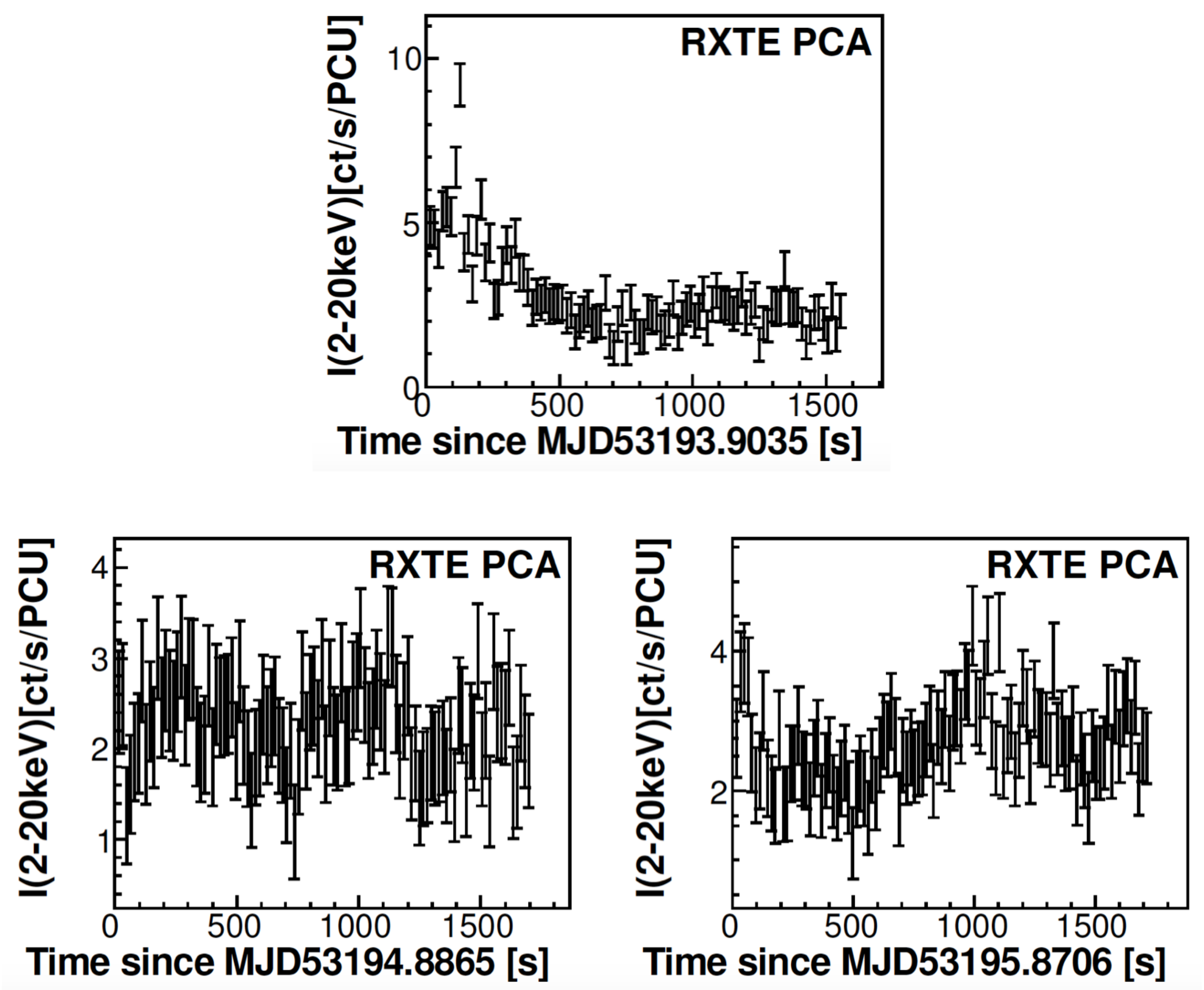}}
  \hfill
  \raisebox{-0.5\height}{\includegraphics[width=0.55\textwidth]{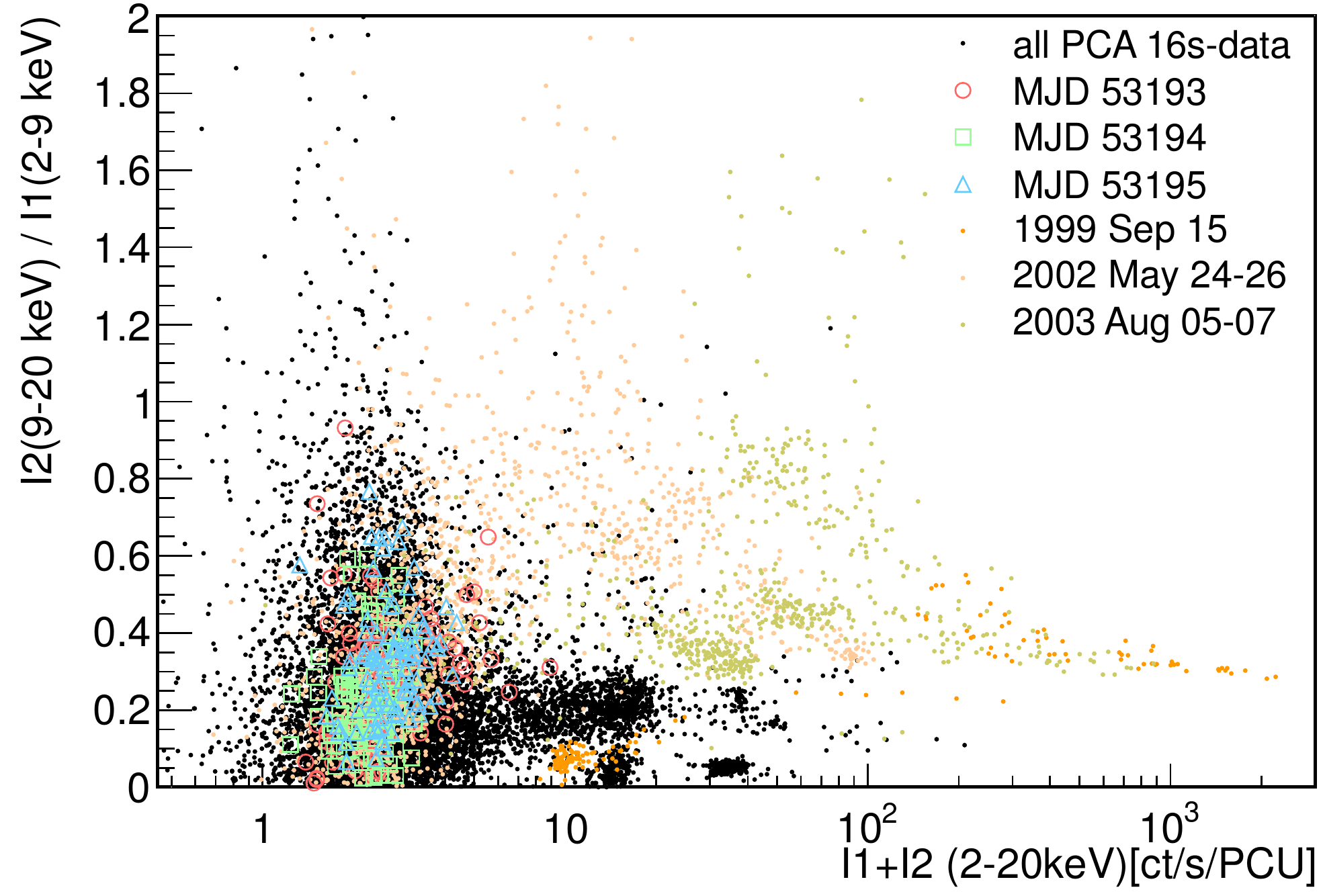}}
  }
   \caption{\emph{RXTE} data on V4641 Sgr showing the PCA lightcurves for the three individual pointings (left plots) and the Hardness-Intensity diagram (right plot) obtained with the \emph{RXTE} PCA data taken during the whole mission of the satellite (black points). Data corresponding to \hess\ observations are highlighted (open symbols) and data corresponding to known flares from the source as observed with \emph{RXTE} data are also highlighted (e.g.~\cite{Wijnands2000}).}
   \label{fig:V4641}
   \end{figure}

The individual PCA lightcurves shown in the left panel of Figure~\ref{fig:V4641} indicate various degrees of X-ray variability with the clearest evidence for flaring visible as a sharp $\sim5$-fold count rate fluctuation during the first observation. In marked contrast, the second observation is uniformly faint with the $\chi^{2}$ probability of constant count rate $P_{\rm const}=0.97$, and hence consistent with a period of steady, low-level emission. Subsequently, the third observation reveals a reemergence of mild variability ($P_{\rm const}=0.07$) with $\sim2$-fold count rate fluctuations occurring on timescales of $\sim500$s.

The HID for V4641 Sgr is shown in the right panel of Figure~\ref{fig:V4641}. Data corresponding to the most famous flaring period of V4641 Sgr, 1999 Sept 15, spanning over 1500s, are displayed for reference (orange markers). Two longer flaring episodes have been highlighted in the HID shown in Figure~\ref{fig:V4641}: 2002, May 24-26 and 2003 Aug 05-07. These episodes show a topology similar to the 1999 episode in the HID diagram, shifted towards lower intensity, but with harder X-rays. Compared to these remarkable events, the observations reported here took place during a much fainter X-ray luminosity. It seems therefore likely that V4641 Sgr underwent a period of only mild activity during our observations.

Technical issues prevented \gr\ data corresponding to the first \emph{RXTE} observation from being obtained. Simultaneous \gr\ observations were obtained corresponding to the second \emph{RXTE} exposure, which showed no indications of X-ray variability. Although the source began to show increased X-ray activity during the third \emph{RXTE} observation, the degree of overlap with the corresponding \hess\ observations was minimal. None of the \hess\ data do not show VHE \gr\ emission. Integral flux upper limits above the instrumental threshold energy corresponding to the overall \hess\ exposure at the position of V4641 Sgr are listed in Table \ref{tab:mqs:hess_uls}.

\begin{table*}[h!]
\caption{\hess\ VHE \gr\ integral flux upper limits above the telescope energy threshold $E_{\rm thresh}$ (anti-correlated with the zenith angle of the observations, $Z$) corresponding to both event selection regimes. The upper limits are derived at the 99\% confidence level, assuming a power law spectrum ($dN/dE\propto E^{-\Gamma}$) with the photon index $\Gamma_{\rm std} = 2.6$ for standard cuts and $\Gamma_{\rm hard} = 2.0$ for hard cuts.}\label{tab:mqs:hess_uls}
\centering
\begin{tabular}{llllll}
\hline\hline
Target & Cuts & $T_{\rm Live}$ [s] & $\bar{Z}_{\max}$ [$^{\circ}$]  & $E_{\rm thresh}$ [GeV]& $I(>E_{\rm thresh})$ [ph cm$^{-2}$s$^{-1}$] \\
\hline
\multirow{2}{*}{GRS 1915+105} & Standard & 24681 & 40.6 &  562 &$<7.3\times10^{-13}$\\
	& Hard & 24681 & 40.6 & 1101 & $<1.1\times10^{-13}$\\
\hline
\multirow{2}{*}{Cir X-1} & Standard & 19433 & 43.6& 562 &$<1.2\times10^{-12}$ \\
	& Hard & 19433 & 43.6 & 1101 & $<4.2\times10^{-13}$\\
\hline
\multirow{2}{*}{V4641 Sgr} & Standard & 6335 & 8.4 &  237 &$<4.5\times10^{-12}$\\
	& Hard & 6335 & 8.4 & 422 & $<4.8\times10^{-13}$ \\
\hline
\end{tabular}
\end{table*}

\vspace{-4mm}
\section{Outlook}
\vspace{-1mm}
Microquasars continue to be classified as targets of opportunity for IACTs, requiring a rapid response to any external trigger to maximise the likelihood of obtaining a significant detection. These conditions are realized with the commissioning of the H.E.S.S.-28m telescope, which provides the world's largest mirror surface and therefore the lowest energy threshold while at the same time allowing for very rapid follow-up observations \cite{Hofverberg}. To exploit these new opportunities and thanks to increasing understanding of the behavior of microquasars, the trigger schemes for TeV follow-up observations have evolved significantly over the last years. In the future, alternative observational strategies, including continuous monitoring of candidate microquasars in the VHE \gr\ band, may become possible using dedicated subarrays of the forthcoming Cherenkov Telescope Array. 
	 
\vspace{-2mm}
\section{Acknowledgments}\vspace{-1mm}
\setstretch{0.9}
\noindent{\footnotesize The support of the Namibian authorities and of the University of Namibia in facilitating the construction and operation of H.E.S.S. is gratefully acknowledged, as is the support by the German Ministry for Education and Research (BMBF), the Max Planck Society, the German Research Foundation (DFG), the French Ministry for Research, the CNRS-IN2P3 and the Astroparticle Interdisciplinary Programme of the CNRS, the U.K. Science and Technology Facilities Council (STFC), the IPNP of the Charles University, the Czech Science Foundation, the Polish Ministry of Science and Higher Education, the South African Department of Science and Technology and National Research Foundation, and by the University of Namibia. We appreciate the excellent work of the technical support staff in Berlin, Durham, Hamburg, Heidelberg, Palaiseau, Paris, Saclay, and in Namibia in the construction and operation of the equipment.}
\bibliographystyle{JHEP}

\end{document}